\documentclass[11pt]{article}
\usepackage{epsfig,cite}
\usepackage{amssymb,amsmath}
\usepackage{times}
\makeatletter \@addtoreset{equation}{section} \makeatother

\mathchardef\varGamma="0100 \mathchardef\varDelta="0101
\mathchardef\varTheta="0102 \mathchardef\varLambda="0103
\mathchardef\varXi="0104 \mathchardef\varPi="0105
\mathchardef\varSigma="0106 \mathchardef\varUpsilon="0107
\mathchardef\varPhi="0108 \mathchardef\varPsi="0109
\mathchardef\varOmega="010A

\def\bfone{\relax{\rm 1\kern-.35em 1}}

\DeclareFontFamily{U}{rsf}{} \DeclareFontShape{U}{rsf}{m}{n}{
  <5> <6> rsfs5 <7> <8> <9> rsfs7 <10-> rsfs10}{}
\DeclareMathAlphabet\Scr{U}{rsf}{m}{n}

\newcommand{\rU}{\mathrm{U}}
\newcommand{\rUSp}{\mathrm{USp}}
\newcommand{\rSU}{\mathrm{SU}}
\newcommand{\rE}{\mathrm{E}}
\newcommand{\rSO}{\mathrm{SO}}
\newcommand{\rO}{\mathrm{O}}
\newcommand{\rSL}{\mathrm{SL}}

\newcommand{\rGL}{\mathrm{GL}}

\begin{document}
\sloppy \raggedbottom
\setcounter{page}{1}

\newpage
\setcounter{figure}{0}
\setcounter{equation}{0}
\setcounter{footnote}{0}
\setcounter{table}{0}
\setcounter{section}{0}

\begin{titlepage}

\thispagestyle{empty}

\begin{flushright}
\hfill{CERN-PH-TH/2005-261}
\end{flushright}

\vspace{35pt}

\begin{center}{ \LARGE{\bf
Spontaneously broken supergravity: Old and new facts\footnote{To
appear in the proceedings of the Conference on ``Quantum Theory
and Symmetries IV'', Verna, Bulgaria, August 2005. }}}
\vspace{60pt}

{\bf  R. D'Auria$^\bigstar$, S. Ferrara$^\dag$ and M.
Trigiante$^\bigstar $}

\vspace{15pt}

$^\bigstar${\it Dipartimento di Fisica, Politecnico di Torino \\
C.so Duca degli Abruzzi, 24, I-10129 Torino, and\\
Istituto Nazionale di Fisica Nucleare, \\
Sezione di Torino,
Italy}\\[1mm] {E-mail: riccardo.dauria@polito.it,  mario.trigiante@polito.it}

$^\dag$ {\it CERN, Physics Department, CH 1211 Geneva 23,
Switzerland\\ and\\ INFN, Laboratori
Nazionali di Frascati, Italy\\and\\
Department of Physics \& Astronomy, University of California, Los Angeles, CA, USA}\\[1mm] {E-mail: Sergio.Ferrara@cern.ch}

\vspace{50pt}

{ABSTRACT}
\end{center}

\medskip

 We report on some recent investigations of the structure of the
four dimensional gauged supergravity Lagrangian which emerges from
flux and Scherk--Schwarz compactifications in higher dimensions.
Special attention is given to the gauge structure of M--theory
compactified on a  seven torus with 4--form and geometrical (spin
connection) fluxes turned on. A class of vacua, with flat
space--time and described by ``no--scale'' supergravity models, is
analyzed.
\end{titlepage}

\newpage


\section{Introduction}
New massive deformations of ``extended'' supergravity theories
have recently been investigated in the context of flux
compactifications from higher dimensional theories. The latter
correspond to superstring or M--theory vacua with some p--form
field strength turned on along the compactified directions
\cite{ps}-\cite{h}. In a more sophisticated mathematical language
they correspond to fluxes when the p-form is integrated on a
p-cycle in the internal manifold.\par The presence of fluxes
determines indeed a non--trivial scalar potential \cite{ps} in the
effective low--energy supergravity, which defines in some cases
vacua with vanishing cosmological constant (at tree level), in
which spontaneous (partial) supersymmetry breaking may occur and
(some of) the moduli of the internal manifold are fixed. In fact
theories with vanishing cosmological constant are generalized
no--scale models, which were studied long ago in the pure
supergravity context \cite{noscale1,noscale2}. The presence of
fluxes also gives rise in the low--energy supergravity to local
symmetries gauged by vector fields \footnote{In four dimensional
supergravities coupled to linear multiplets, fluxes may give rise
to more general couplings.}. Supergravity models with such gauge
symmetries (gauged supergravities) have been extensively studied
in the literature \cite{dwn}-\cite{dwst1}, also in connection to
flux compactifications or Scherk--Schwarz dimensional reduction
\cite{ss},\cite{css}-\cite{vz}.  Actually in extended
supergravities (${ N}\ge 2$) the \emph{gauging} procedure, which
consists in promoting a global symmetry group of the Lagrangian to
local invariance, is the only way for introducing a non--trivial
scalar potential without explicitly breaking supersymmetry. The
global symmetry group of extended supergravities is the isometry
group $G$ of the scalar manifold, whose non--linear action on the
scalar fields is associated with an electric/magnetic duality
action on the $n_v$ vector field strengths and their duals
\cite{gz}. This duality transformation is required in four
dimensions to be symplectic and thus is defined by the embedding
of $G$ inside ${\rm Sp}(2n_v,\mathbb{R})$. Gauge symmetries
deriving from flux compactifications typically are related to
non--semisimple Lie groups ${\Scr G}$ containing abelian
translational isometries acting on axionic fields which originate
from ten dimensional R--R forms $C_{(p)}$ ($p=0,2,4$ for Type IIB)
or the NS 2--form $B_{(2)}$. The embedding of ${\Scr G}$ inside
$G$ is defined at the level of the corresponding Lie algebras by
the flux tensors themselves, which play the mathematical role of
an \emph{embedding matrix} \cite{dwst1}. \par No--scale models
arising from flux compactifications or Scherk--Schwarz dimensional
reduction give rise to a semi--positive definite scalar potential
which has an interpretation in terms of an ${ N}$--extended gauged
supergravity in four dimensions. Let us recall the general form of
such scalar potential $V(\Phi)$, $\Phi$ denoting collectively the
scalar fields, \cite{fm,cgp,df}:
\begin{eqnarray}\delta_B^A V(\Phi)=-3 S^{AC}S_{BC}+N^{IA} N_{IB}\,,\label{potential}\end{eqnarray}
where $S_{AB}=S_{BA}$,  and $N^{IA}$ appear in the gravitino and
spin $1/2$ supersymmetry transformations
 \begin{eqnarray} \delta\psi_{A\mu}&=&\frac 1 2
S_{AB}\gamma_\mu\epsilon^B+\cdots\noindent\\
 \delta\lambda^I&=&
N^{IA}\epsilon_A+\cdots\,,\label{varfer}\end{eqnarray} and give
rise  in the supergravity Lagrangian to the following terms:
\begin{eqnarray}\frac{1}{\sqrt{-g}}\mathcal{L}&=&\cdots
+S_{AB}\bar\psi_\mu^A\sigma^{\mu\nu}\psi_\nu^B+iN^{IA}\bar\lambda_I\gamma^\mu\psi_{\mu
A}  -\!V(\Phi)\,.\end{eqnarray}
 Flat space demands that on the extremes ${\partial V}/{\partial\Phi}=0$ the potential  vanishes, so
\begin{eqnarray}3\sum_CS^{AC}S_{CA}&=&\sum_IN^{IA} N_{IA},\qquad \forall A\,,\end{eqnarray}
  The first term in the potential
(\ref{potential}) is the square of the gravitino mass matrix. It
is hermitian, so it can be diagonalized by a unitary
transformation. Assume that it is already diagonal, then the
eigenvalue in the entry $(A_0,A_0)$ is non zero if and only if
$N^{IA_0}\neq 0$ for some $I$. On the other hand, if the gravitino
mass matrix vanishes then $N^{IA}$ must be zero.

 For no-scale models \cite{noscale1,noscale2}, there is a subset of fields $\lambda^{I'}$ for which
 \begin{eqnarray}3\sum_CS^{AC}S_{CA}&=&\sum_{I'}N^{I'A} N_{I'A},\qquad \forall
A\label{cancel}\end{eqnarray} at any point in the scalar manifold
${\Scr M}_{scal}$. This implies that the potential is given by
 \begin{eqnarray}V(\Phi)&=&\sum_{I\neq I'}N^{IA} N_{IA}\,,\end{eqnarray}
 and it is manifestly positive definite. Zero vacuum energy on a point of ${\Scr M}_{scal}$ implies that
$N^{IA}=0$, $I\neq I'$  at that point. This happens independently
of the number of unbroken supersymmetries, which is controlled by
$N^{I'A}$.\par
 In the sequel we shall discuss no--scale models as they originate
 from M--theory compactifications on a twisted seven--torus with 4--form flux, 7--form flux
 and geometrical flux \cite{km}-\cite{h},\cite{ddf}-\cite{dft4}.\par
 A twisted torus corresponds, in this framework, to the so called
 Scherk--Schwarz compactification, i.e. to the replacement of a
 flat torus $T^7$ with a seven--dimensional group--manifold  whose
 structure constants $\tau_{IJ}{}^K$ (from now on the capital latin indices label
 the seven directions of the internal torus: $I,J,M,N...=1,\dots, 7$) determine the Lie algebra of
 the ``graviphoton fields'' $A^I_\mu$ associated with the
 Kaluza--Klein mixed components of the metric $g_{I\mu}$. The
 corresponding four dimensional curvatures are therefore:
\begin{eqnarray}
F^I&=&dA^I+\frac{1}{2}\,\tau_{KL}{}^I\, A^K\wedge A^L\,.
\end{eqnarray}
The internal curvature of the eleven dimensional 3--form field
$C^{(3)}$ is given by:
\begin{eqnarray}
F^{(0)}_{IJKL}&=&-g_{IJKL}-\frac{3}{2}\tau_{[IJ}{}^M\,C_{KL]M}\,,\label{eq15}
\end{eqnarray}
while the external (space--time)  components of the same field
strength read:
\begin{eqnarray}
F^{(4)}&=&d{\cal A}^{(3)}- g_{IJKL}\,A^I\wedge A^J\wedge A^K\wedge
A^L- B_I\wedge F^I\,,\label{eq6}
\end{eqnarray}
where ${\cal A}^{(3)}$ denote the (non--propagating) four
dimensional 3--form field and $B_{\mu\nu I}$ are the seven
antisymmetric tensor fields originating from the dimensional
reduction of $C^{(3)}$. The constants $g_{IJKL},\, \tau_{IJ}{}^K$
are bounded to satisfy the following relations:
\begin{eqnarray}
\tau_{IJ}{}^J&=&0\,\,;\,\,\,\tau_{[IJ}{}^M\,\tau_{K]M}{}^L=0\,\,;\,\,\,\,\tau_{[IJ}{}^P\,g_{KLM]P}=0\,.
\end{eqnarray}
These constraints ensure that, when massive antisymmetric tensors
are suitably dualized to massive vector fields, so that the
M--theory in $D=4$ admits a global (on-shell) $\rE_{7(7)}$
symmetry, a 28--dimensional Lie algebra is gauged, whose structure
constants are given in terms of $g_{IJKL},\, \tau_{IJ}{}^K$ and
$\tilde{g}$ (where $\tilde{g}$ is the flux associated with the
space-time components of the 4--form:
$F^{(4)}_{\mu\nu\rho\sigma}\propto
\tilde{g}\,\epsilon_{\mu\nu\rho\sigma}$). In section 2 we discuss
the equations of motion and the potential of M--theory
compactification on a twisted torus with internal fluxes turned
on. In section 3 we discuss flat vacua and the correspondence to
the Scherk--Schwarz breaking. In section 4 we discuss these
results in terms of the gauging of a subalgebra of $\rE_{7(7)}$.
We refer the reader to the appendix for a description of the dual
gauge algebra as a subalgebra of $\rE_{7(7)}$.

\section{The equations of motion and the potential}
The bosonic equations of motion of M--theory can be obtained by
varying the Lagrangian with respect to the vielbein 1--form $V^a$
and the 3--form $C^{(3)}$. \par The $g_{\mu\nu}$, $G_{IJ}$ and
$A^I$ field equations come from
 the eleven dimensional Einstein equations:
{\small {\small\begin{eqnarray}\label{einst}
R_{\mu\nu}-\frac{1}{2}\,g_{\mu\nu}\,R&=&T_{\mu\nu}\,,\nonumber\\
R_{\mu I}+\frac{1}{2}\,g_{\mu I}\,R&=&T_{\mu I}\,,\nonumber\\
R_{IJ}+\frac{1}{2}\,G_{IJ}\,R&=&T_{IJ}\,,
\end{eqnarray}}}
where $g_{\mu I}=G_{IJ}\,A^J_\mu$ and $G_{IJ}$ are the coordinates
of ${\rGL}(7)/{\rSO(7)}$\footnote{In our notations $G_{IJ}$ is a
positive definite matrix and we adopt the ``mostly minus''
convention for the space--time metric.}. The tensor $T$ is the
energy momentum tensor of the 4--form. Incidentally we remark that
in this formulation the $R$--symmetry
 of the corresponding $N=8$ supergravity is $Spin(7)$, the eleven dimensional
gravitino gives rise to eight gravitinos which are in the
eight--dimensional spinorial representation and to spin $1/2$
which transform in the ${\bf 8}+{\bf 48}$ of the same group.\par
 The 3--form field equations read as follows:\footnote{For the eleven dimensional equations we are using the conventions
 and notations of reference \cite{fd}}
{\small\begin{eqnarray} d \star F^{(4)}&=& \frac{1}{4}\,
F^{(4)}\wedge F^{(4)}\,.\label{gff}
\end{eqnarray}}
Since in this paper we are mainly concerned with the general form
of the scalar potential coming from the twist and the fluxes, we
will carefully analyze this equation only for those entries which
receive contributions from the scalar potential \cite{dft3}. Let
us write the dual of the field equations originating from the
Euler--Lagrange equations for $\mathcal{A}_{\mu\nu\rho}$ and
$C_{IJK}$.  The first equation  allows us to integrate out the
$\mathcal{A}_{\mu\nu\rho}$ field in a manner which we shall
explain in a moment. This integration gives an extra contribution
to the scalar potential coming from the Chern--Simons term. The
second equation contains the derivative of the vacuum energy with
respect to $C_{IJK}$ and contributes to the equation of motion of
the $C_{IJK}$ scalar.\par Let us define the following 4--D scalar
quantity: {\small\begin{eqnarray}
 P&=& \frac{1}{\sqrt{-g}}\,\epsilon^{\mu_1\dots \mu_4}\, F^{(4)}_{\mu_1\dots \mu_4}\,,\label{P}
\end{eqnarray}}
where  $F^{(4)}_{\mu_1\dots \mu_4}$ was defined in eq.
(\ref{eq6}). For the purpose of computing the scalar potential,
 only the $d {\cal A}^{(3)}$ part of $F^{(4)}$ will be relevant.
The $\mathcal{A}_{\mu\nu\rho}$ field equation then reads:
{\small\begin{eqnarray}
 d (V_7\,P)&=&-\frac{1}{4}\, F^{(1)}_{IJK}\, F^{(0)}_{PQRS}\,\epsilon^{IJKPQRS}\,,\label{dP}
\end{eqnarray}}
where the field strength $F^{(1)}_{IJK}$ is defined as follows
\cite{dft3}:
\begin{eqnarray}
F^{(1)}_{IJK}&\equiv&{\Scr
D}^{(\tau)}C_{IJK}-\tau_{[IJ}{}^L\,A_{K]L}+4\,g_{IJKL} \,A^L\,,
\end{eqnarray}
the covariant derivative ${\Scr D}^{(\tau)}$ corresponding to the
gauge connection defined by $\tau_{IJ}{}^K$ and $A_{IJ}$ being the
21 vector fields originating from $C^{(3)}$.
 For our purposes we shall also restrict ourselves to
the ${\Scr D}^{(\tau)}C_{IJK}$ term in $F^{(1)}_{IJK}$. Equation
(\ref{dP}) implies that its right hand side is a closed form. In
fact the crucial ingredient is that the term $F^{(1)}_{IJK}\,
F^{(0)}_{PQRS}\,\epsilon^{IJKPQRS}$ is an exact form on the
twisted torus with fluxes, and it can be written as
{\small\begin{eqnarray}
 F^{(1)}_{IJK}\, F^{(0)}_{PQRS}\,\epsilon^{IJKPQRS}&=&-d\left(C_{IJK}\,
 (g_{PQRS}+\frac{3}{4}\,\tau_{[PQ}^N\,C_{RS]N})\epsilon^{IJKPQRS}+\tilde{g}\right)\,,
\end{eqnarray}}
where the integration constant $\tilde{g}$ \cite{d} is actually
related to the dual gauge algebra in the $\rE_{7(7)}$ covariant
formulation described in \cite{dft2}. From this we get the value
of $V_7\, P$ to be: {\small\begin{eqnarray} V_7\,
P&=&\frac{1}{4}\,\left(
C_{IJK}\,(g_{LPQR}+\frac{3}{4}\,\tau_{[LP}^N\,C_{QR]N})\,\epsilon^{IJKLPQR}+\tilde{g}\right)\,.\label{dP1}
\end{eqnarray}}
Note the important identity: {\small\begin{eqnarray}
V_7\,\frac{\delta P}{\delta
C_{IJK}}&=&-\frac{1}{4}\,\epsilon^{IJKLPQR}\,
F^{(0)}_{LPQR}\,.\label{dP2}
\end{eqnarray}}
Let us now turn to considering the equation of motion for the
$C_{IJK}$ fields. They read: {\small\begin{eqnarray}
\partial_\mu\left(V_7\,\sqrt{-g} G^{I_1 J_1}G^{I_2 J_2}G^{I_3 J_3}\,
g^{\mu\nu}\, \partial_\nu
C_{J_1J_2J_3}\right)&=&-\frac{3}{2}\,\frac{1}{7!}\,\epsilon^{\mu\nu\rho\lambda}\,
F_{IJKP}\, F_{\mu\nu\rho\lambda}\, \epsilon^{I_1I_2I_3
IJKP}+\nonumber\\&&-\frac{1}{2}\,V_7\,\sqrt{-g}\,
\tau_{PQ}{}^{[I_1}\, F^{I_2I_3]PQ}\,.\label{dC}
\end{eqnarray}}
By using equations (\ref{dP1}) and (\ref{dP2}) and the fact that:
{\small\begin{eqnarray} \frac{\delta (
F^{(0)}_{IJKL}\,F^{(0)\,IJKL})}{\delta C_{PQR}}&=&-3\,
\tau_{IJ}^{[P}\, F^{(0)QR]IJ}\,,\label{delta2}
\end{eqnarray}}
equation (\ref{dC}) can be rewritten in the form:
{\small\begin{eqnarray}
\partial_\mu\left(V_7\,\sqrt{-g} G^{I_1 J_1}G^{I_2 J_2}G^{I_3 J_3}\,
g^{\mu\nu}\, \partial_\nu
C_{J_1J_2J_3}\right)&=&\sqrt{-g}\,\frac{\delta V}{\delta
C_{I_1I_2I_3}}\,,
\end{eqnarray}}
where the $C_{IJK}$--dependent part of the potential is:
{\small\begin{eqnarray}
V_C &=& \frac{3}{16}\,\frac{1}{7!}\,\frac{1}{V_7}\, \left( C_{IJK}\,(g_{LPQR}+\frac{3}{4}\,\tau_{[LP}^N\,C_{QR]N})\,\epsilon^{IJKLPQR}+\tilde{g}\right)^2+\nonumber\\
&&+\frac{1}{6}\,V_7\,F^{(0)}_{IJKL}\,F^{(0)}_{MNPQ}\, G^{IM}\,
G^{JN}\, G^{KP}\, G^{LQ}\,,\label{vc}
\end{eqnarray}}
where $F^{(0)}_{IJKL}$ is given in eq. (\ref{eq15}). One can
easily compute the scalar potential in the Einstein frame by
noting that {\small\begin{eqnarray} g_{\mu\nu}&=&\frac{1}{V_7^2}\,
g_{\mu\nu}^{E}\,.
\end{eqnarray}}
Therefore in this frame, the potential becomes multiplied by an
overall $(V_7)^{-2}$.\par The full scalar potential in the
Einstein frame is thus obtained by adding to $V_C$ the
Scherk--Schwarz purely $G$--dependent part originating from the
eleven--dimensional Einstein term. It is useful to write the
entire potential as the following sum: {\small\begin{eqnarray}
V&=&V_E+V_K+V_{C-S}\,,\label{v}
\end{eqnarray}}
 where the three terms on the right hand side originate from the eleven dimensional Einsten, kinetic and Chern--Simons terms respectively, and are found to have the following expression:
{\small\begin{eqnarray}
V_E&=&\frac{1}{V_7}\,\left(2\, G^{KL}\, \tau_{KJ}{}^I\, \tau_{LI}{}^J+G_{II'}\,G^{JJ'}\, G^{KK'}\,\tau_{JK}{}^I\,\tau_{J'K'}{}^{I'}\right)\,,\nonumber\\
V_K&=&\frac{3}{16}\,\frac{1}{7!}\,\frac{1}{V_7}\,(g_{IJKL}+\frac{3}{2}\,\tau_{[IJ}^R\,C_{KL]R})(g_{MNPQ}+
\frac{3}{2}\,\tau_{[MN}^R\,C_{PQ]R})\, G^{IM}\, G^{JN}\, G^{KP}\, G^{LQ}\,,\nonumber\\
V_{C-S}&=& \frac{1}{6}\,\frac{1}{V_7^3}\, \left(
C_{IJK}\,(g_{LPQR}+\frac{3}{4}\,\tau_{[LP}^N\,C_{QR]N})\,\epsilon^{IJKLPQR}+\tilde{g}\right)^2\,.\label{vekcs}
\end{eqnarray}}
Recall that in our conventions $G_{IJ}$ is a positive definite
matrix. Note that for $\tau=g=0$ we just get a positive
cosmological constant, as noted in \cite{d}.
 \section{ Flat group vacua of the potential}
The scalar potential in (\ref{v}) and (\ref{vekcs}) has the
property that $V_K\ge 0$, $V_{C-S}\ge 0$ while $V_E$ has no
definite sign \cite{ss}. Therefore in general we may have vacua
with different signs of the cosmological constant.\par From
inspection of the scalar potential, let us make some general
comments on the possible vacua of this class of models. We start
analyzing the equation $\delta V/\delta C_{IJK}=0$, necessary in
order to have a bosonic background with $C_{IJK}\equiv
C^{0}_{IJK}=constant$. Because of the properties (\ref{dP2}) and
(\ref{delta2}) this is ensured by setting $F_{IJKL}^{(0)}=0$ or
equivalently: {\small\begin{eqnarray} g_{IJKL}+\frac{3}{2}\,
\tau_{[IJ}{}^P\, C_{KL]P}&=&0\,.\label{f40}
\end{eqnarray}}
As a consequence of equation (\ref{f40}), \emph{the 4--form flux
has always a vanishing contribution ($V_K=0$) to the vacuum
energy}. Next we extremize the potential with respect to the
volume of the torus $V_7$. Taking into account the dependence of
the internal metric $G_{IJ}$ on $V_7$, given by:
\begin{eqnarray}
G_{IJ}&=& (V_7)^{\frac{2}{7}}\,
\hat{G}_{IJ}\,\,\,\,;\,\,\,\,\,\,{\rm det}(\hat{G})=1\,.
\end{eqnarray}
and using the following short-hand notation:
\begin{eqnarray}
b&=&2\, G^{KL}\, \tau_{KJ}{}^I\, \tau_{LI}{}^J+G_{II'}\,G^{JJ'}\,
G^{KK'}\,\tau_{JK}{}^I\,\tau_{J'K'}{}^{I'}\,,\nonumber\\
a&=&
C_{IJK}\,(g_{LPQR}+\frac{3}{4}\,\tau_{[LP}^N\,C_{QR]N})\,\epsilon^{IJKLPQR}+\tilde{g}\,,\nonumber
\end{eqnarray}
the two conditions $\delta V/\delta C_{IJK}=0=\delta V/\delta V_7$
will reduce the expression of the potential at the minimum $V_0$
to:
\begin{eqnarray}
V_0&=& -\frac{4}{3}\,\frac{a^2}{(V_7^0)^3}\le
0\,\,\,\mbox{where}\,\,\,\,\,b\,(V_7^0)^{\frac{12}{7}}=-\frac{7}{3}\,a^2\,.\label{vo}
\end{eqnarray}
From the above equations we conclude that a necessary condition
for a vacuum to exist is $V_E\le 0$ and that at the minimum $V\le
0$. This excludes the existence of a de Sitter vacuum (i.e.
maximally symmetric space-time geometry with positive cosmological
constant).\par
 A
particular appealing class of models, which correspond to
``no--scale'' supergravities \cite{noscale1,noscale2},
 are obtained for those gaugings for which $V=0$. This defines a ``flat group'' \cite{ss}.
From equations (\ref{vo}) this implies that $a=b=0$, namely that
$V_E=V_{C-S}=0$, and that $V_7$ is an unfixed modulus. Condition
$V_{C-S}=0$ in turn implies: {\small\begin{eqnarray}
P&=&0\,\,\,\,\Leftrightarrow\,\,\,\,\,C_{MNR}\,\left(g_{IJKL}+\frac{3}{4}\,
\tau_{[IJ}{}^P\,
C_{KL]P}\right)\,\epsilon^{MNRIJKL}+\tilde{g}=0\,.\label{P0}
\end{eqnarray}}
The second equation can also be written as the following condition
on $\tilde{g}$: {\small\begin{eqnarray} \tilde{g}&=&\frac{3}{4}\,
C^0_{MNR}\,\tau_{[IJ}{}^P\,
C^0_{KL]P}\,\epsilon^{MNRIJKL}\,,\label{tildeg}
\end{eqnarray}}
where $C^0_{IJK}$ is a solution of equation (\ref{f40}) and thus
depends on $g_{IJKL}$. This equation ensures that the $G_{IJ}$
moduli equations are the same as in the $g=\tilde{g}=0$ case,
because the $F$--contribution to the energy--momentum tensor
vanishes in these vacua. Condition $V_E=0$ on the other hand
 implies restrictions
  of the $\tau$ matrices. These were described in the pioneering paper of ref. \cite{ss} for $g_{IJKL}=\tilde{g}=0$. \par
Summarizing, a necessary condition for our models to admit
Minkowski vacua is that the form-fluxes $g_{IJKL}$ and $\tilde{g}$
satisfy, besides $V_K=0$ also $V_{C-S}=0$. If we associate the
background quantities  $\tau_{IJ}{}^K,\,g_{IJKL}$ and $\tilde{g}$
 with components of a larger
representation of the group $\rE_{7(7)}$, it can be shown that
conditions $V_K=0=V_{C-S}$ amounts to stating that $g_{IJKL}$ and
$\tilde{g}$ can be generated by acting on $\tau_{IJ}{}^K$ by means
of an $\rE_{7(7)}$ transformation or, equivalently, that all the
models admitting Minkowski vacua belong to the same
$\rE_{7(7)}$--orbit as the model with $g_{IJKL}=\tilde{g}=0$
originally considered by Scherk and Schwarz, and thus share with
it the same physics (mass spectrum etc...). Therefore there is an
underlying hidden $\rE_{7(7)}$ symmetry which is not manifest in
the formulation of these models with tensor fields, but which is
apparent at the level of equations of motion and Bianchi
identities in the dual description of this compactification in
which the antisymmetric tensor fields are replaced by scalar
fields. This global symmetry however holds only if, besides the
fields, the background quantities are transformed as well, and
thus should not be regarded as a symmetry of the theory, but
rather as a mapping between two different theories (a proper
duality). This justifies \emph{a posteriori} the aforementioned
identification of the background quantities
$\tau_{IJ}{}^K,\,g_{IJKL}$ and $\tilde{g}$ with parts of an
$\rE_{7(7)}$ representation.
\par
 To make a concrete example, let us
consider the case in which $I=0,i$, $i=1,\dots, 6 $ with
$\tau_{IJ}{}^K=\tau_{0i}{}^j$, zero otherwise, and
$g_{IJKL}=g_{0ijk}$, zero otherwise. In this case
$\tau_{0i}{}^j=T_j{}^i$ is chosen to be an antisymmetric matrix of
rank 3 which can be set in the form: {\small\begin{eqnarray}
T_i{}^j&=&\left(\begin{matrix}m_1\,\epsilon &0&0\cr
0&m_2\,\epsilon &0\cr 0&0&m_3\,
\epsilon\end{matrix}\right)\,\,\,;\,\,\,\,\epsilon=\left(\begin{matrix}0&1\cr
-1&0\end{matrix}\right)\,.
\end{eqnarray}}
In this context  the equation (\ref{f40}) becomes
$F^{(0)}_{0ijk}=0$ which fixes all $C_{ijk}$ fields but not the
$C_{0ij}$ scalars. The $C_{0ij}$ fields give masses to the
$A_{ij}$ vector fields with the exception of the three entries
$(ij)=(1,2),(3,4),(5,6)$. Therefore three of the $C_{0ij}$ scalar
remain massless moduli. The $G_{IJ}$--sector gives, as discussed
in reference \cite{ss}, four additional massless scalars, of which
two are the volume $V_7$ and $G_{00}$ and two other come from
internal components of the metric.\par If one further discusses
the spectrum of the remaining fields, the six vectors $A_{i0}$ are
eaten by the six antisymmetric tensors $B_i$ because of the
magnetic mass term in the free differential algebra \cite{dft3}.
An additional massless scalar comes from the massless 2--form
$B_0$ and finally an additional massless vector comes from the
$A^0$ Kaluza--Klein vector. The other six $A^i$ vectors become
massive because of the twisting of the torus. We conclude that in
this theory there are always eight massless scalars and four
massless vectors, in agreement with \cite{ss}. The effect of
turning on $g$ and $\tilde{g}$ is not of giving extra masses, but
of shifting the v.e.v. of the $C_{IJK}$ fields. This can be
understood by an extension of the flat group where $g$ and
$\tilde{g}$ play the role of additional structure constants. In
the next section we will recover this result as well as the form
of the potential, from the underlying duality symmetry of the dual
formulation of the theory, in which all antisymmetric tensors
$B_I$ are dualized into scalars $\tilde{B}^I$ and the $\rE_{7(7)}$
symmetry is recovered.

\section{The dual gauge algebra and its scalar potential}
We now interpret the above result in the usual formulation of the
four dimensional theory based on the flat gauging
\cite{ss,css,svn,adfl0}. From the results of \cite{dft2} this
amounts to dualizing those vector fields which participate to the
anti--Higgs mechanism, in our case they are the $A_{0i}$ 1--forms,
which are therefore replaced by their $A^{0i}$ magnetic duals. The
dual gauge algebra therefore contains the following 28 generators:
{\small\begin{eqnarray} W^{ij},\, W_{i},\,Z_i,\,Z_0\,.
\end{eqnarray}}
with structure constants obtained from eq. (2.13) of \cite{dft2}.
The first 27 generators form an abelian algebra, and the only non
vanishing commutators are those involving $Z_0$ and given by:
{\small\begin{eqnarray}
\left[Z_0,\,Z_i\right]&=&T_j{}^j\,Z_j-12\,g_{0ijk}\, W^{jk}+\tilde{g}\, W_i\,\nonumber\\
\left[Z_0,\,W^{pq}\right]&=&2\, T_i{}^{[p}\,W^{q]i}-12\,g_{0ijk}\, \epsilon^{ijkpql}\, W_l\,\nonumber\\
\left[Z_0,\,W_i\right]&=& T_j{}^j\,W_j\,,\label{gaugealg}
\end{eqnarray}}
where with respect to \cite{dft2} the redefinition $g\rightarrow
-12\,g$ was made. This algebra defines a flat subalgebra of
$\rE_{7(7)}$ which fits the class of models discussed by Cremmer,
Scherk and Schwarz in \cite{css} and in \cite{svn}, as it was
shown in \cite{alt} and in \cite{dft2}. The gauged supergravity
interpretation was given in \cite{adfl} and the corresponding
gauge algebra is the semidirect product of a $\rU(1)$ by a
27--dimensional abelian algebra and is contained in the branching
of $\rE_{7(7)}$ with respect to $\rE_{6(6)}\times \rO(1,1)$:
{\small\begin{eqnarray} {\bf 133}&\rightarrow & {\bf 1}_0+{\bf
78}_0+{\bf 27'}_{+2}+{\bf 27}_{-2}\,.
\end{eqnarray}}
To compare with the geometrical twist we further branch
$\rE_{6(6)}$ with respect to $\rSL(6)\times \rSL(2)$:
{\small\begin{eqnarray}
{\bf 78}&\rightarrow & ({\bf 35},{\bf 1})+({\bf 1},{\bf 3})+({\bf 20},{\bf 2})\,,\label{branch78}\\
{\bf 27}&\rightarrow & ({\bf 15}',{\bf 1})+({\bf 6},{\bf 2})\,.
\end{eqnarray}}
Our gauging corresponds to the following choice of the ``twist
matrix'' (see \cite{alt} and equation (2.9) of \cite{dft2}):
{\small\begin{eqnarray} Z_0&=&-\frac{2}{3}\,T_i{}^j\,
t_j{}^i+\,g_{0ijk}\, t^{ijk}+\frac{1}{9}\,\tilde{g}\, t_0\,,
\end{eqnarray}}
where we have used the notations introduced in \cite{dft2}. Here
$t_i{}^j$ are the generators of the maximal compact subgroup of
$\rSL(6)$, namely $\rSO(6)$, while $t^{ijk}$ and $t_0$ are
nilpotent generators: the former belong to the $({\bf 20},{\bf
2})$ representation in (\ref{branch78}) with positive grading with
respect to the $\mathfrak{o}(1,1)$ generator of $\rSL(2)$ and the
latter is the nilpotent generator of  $\rSL(2)$ with positive
grading with respect to the same generator. In the same framework
we now discuss the form of the scalar potential, which is expected
not to depend on the dualization procedure. In the dual
formulation this potential is given by \cite{svn,adfl,dgftv}:
{\small\begin{eqnarray} V&=&e^{-6\,\phi}\,\left(
\frac{1}{2}\,(P_{0\,\hat{i}}{}^{\hat{j}})+\frac{1}{6}\,(P_{0\,\hat{i}\hat{j}\hat{k}})^2+(P_0{}^0)^2\right)=V_E+V_K+V_{C-S}\,,
\end{eqnarray}}
where  $\phi$ is the modulus associated with the $0^{th}$ internal
dimension of compactification, the hatted indices are rigid
$\rSO(6)$ indices, while the quantity $P_0$ has value  in the
42--dimensional non--compact part of the $\mathfrak{e}_{6(6)}$ Lie
algebra and represents the vielbein of the five--dimensional
$\sigma$--model $\rE_{6(6)}/\rUSp(8)$.  It is defined as follows:
{\small\begin{eqnarray} P_0=(L^{-1}\, Z_0\,
L)_{|\mbox{non--compact}}\,,
\end{eqnarray}}
where $L$ is the five--dimensional coset representative which,
using the solvable Lie algebra parametrization of
$\rE_{6(6)}/\rUSp(8)$, can be directly written in terms of our
scalar fields as follows: {\small\begin{eqnarray} L&=&
e^{\tilde{B}^0\, t_0}\, e^{\frac{1}{6}\,C_{ijk}\, t^{ijk}}\,
\mathbb{E}\,\,\,;\,\,\,\,\mathbb{E}\in \frac{\rGL(6)}{\rSO(6)}\,.
\end{eqnarray}}
Direct computation shows that: {\small\begin{eqnarray}
P_{0\,\hat{i}\hat{j}}&=&T_i{}^j\, \mathbb{E}^{-1}_{(\hat{i}}{}^i\, \mathbb{E}_{j|\hat{j})}\,,\nonumber\\
P_{0\,\hat{i}\hat{j}\hat{k}}&\propto &(g_{0ijk}+\frac{3}{4}\,T_{[i}{}^n\,C_{jk]n})\,\mathbb{E}^{-1}_{\hat{i}}{}^i\,\mathbb{E}^{-1}_{\hat{j}}{}^j\,\mathbb{E}^{-1}_{\hat{k}}{}^k\,,\nonumber\\
P_{0}{}^0&\propto & \epsilon^{lmnijk}\,
C_{lmn}\,(g_{0ijk}+\frac{3}{8}\,T_{[i}{}^n\,C_{jk]n})+ \tilde{g}
\,.\label{ps}
\end{eqnarray}}
In this language the eight massless modes come from $\tilde{B}^0$,
three from $C_{0ij}$, one from $\phi$ and three from the metric
$G_{ij}$. The latter  can be understood from the fact that under
$\rSO(6)$ these moduli transform in the ${\bf 1}+{\bf 20}^\prime$
and the ${\bf 20}^\prime$ has two vanishing weights.

\section*{Acknowledgments}

 Work supported in part by the European
Community's Human Potential Program under contract
MRTN-CT-2004-005104 `Constituents, fundamental forces and
symmetries of the universe', in which R. D'A. and M.T.  are
associated to Torino University. The work of S.F. has been
supported in part by European Community's Human Potential Program
under contract MRTN-CT-2004-005104 `Constituents, fundamental
forces and symmetries of the universe', in association with INFN
Frascati National Laboratories and by D.O.E. grant
DE-FG03-91ER40662, Task C.

\section*{Appendix}
We can  consider the $\rE_{7(7)}$ generators in the
$\rGL(7,\mathbb{R})$--basis. This corresponds to the branching:
\begin{eqnarray}
{\bf 133}&\rightarrow & {\bf 48}_0+{\bf 1}_0+{\bf
35}_{+2}+\overline{{\bf 35}}_{-2}+{\bf 7 }_{-4}+\overline{{\bf
7}}_{+4}\,.\nonumber
\end{eqnarray}
The coset $\rE_{7(7)}/\rSU(8)$ can be parametrized as follows:
\begin{eqnarray}
\frac{\rE_{7(7)}}{\rSU(8)}&\equiv &
\frac{\rGL(7,\mathbb{R})}{\rSO(7)}\ltimes\,\mbox{Span}({\bf
35}_{+2}+\overline{{\bf 7}}_{+4})\,.\nonumber
\end{eqnarray}
In this basis the $\rE_{7(7)}$ generators are:
\begin{eqnarray}
t_M{}^N&\in& \mathfrak{gl}(7)\,,\nonumber\\
t^{MNP},\,t_{MNP},\,t_P,\,t^P&\in &{\bf 35}_{+2}+\overline{{\bf
35}}_{-2}+\overline{{\bf 7}}_{+4}+{\bf 7 }_{-4}\,.\nonumber
\end{eqnarray}
\begin{eqnarray}
\left[t_M{}^N,\,t_P{}^Q\right]&=&\delta_{P}^N\,t_{M}{}^Q-\delta_{M}^Q\,t_{P}{}^N\,,\nonumber\\
\left[t_M{}^N,\,t^{P_1P_2P_3}\right]&=&-3\,\delta_M^{[P_1}\,t^{P_2P_3]N}+\frac{5}{7}\,\delta_{M}^N\,t^{P_1P_2P_3}\,,\nonumber\\
\left[t_M{}^N,\,t_{P}\right]&=&\delta_P^{N}\,t_{M}+\frac{3}{7}\,\delta_{M}^N\,t_{P}\,,\nonumber\\
\left[t^{N_1N_2N_3},\,t^{P_1P_2P_3}\right]&=&\epsilon^{N_1N_2N_3P_1P_2P_3
Q }\,t_Q\,,\nonumber\\
\left[t_M{}^N,\,t_{P_1P_2P_3}\right]&=&3\,\delta_{[P_1}^{N}\,t_{P_2P_3]M}-\frac{5}{7}\,\delta_{M}^N\,t_{P_1P_2P_3}\,,\nonumber\\
\left[t_M{}^N,\,t^{P}\right]&=&-\delta_M^{P}\,t^{N}-\frac{3}{7}\,\delta_{M}^N\,t^{P}\,,\nonumber\\
\left[t_{N_1N_2N_3},\,t_{P_1P_2P_3}\right]&=&\epsilon_{N_1N_2N_3P_1P_2P_3
Q }\,t^Q\,,\nonumber\\
\left[t^N,\,t_{M}\right]&=&t_M{}^N+\frac{1}{7}\,\delta_M^N\,t\,,\nonumber\\
\left[t^M,\,t^{N_1N_2N_3}\right]&=&-\frac{1}{6}\,\epsilon^{M
N_1N_2N_3 P_1P_2P_3}\,t_{P_1P_2P_3}\,,\nonumber\\
\left[t_M,\,t_{N_1N_2N_3}\right]&=&-\frac{1}{6}\,\epsilon_{M
N_1N_2N_3 P_1P_2P_3}\,t^{P_1P_2P_3}\,\nonumber\\
\left[t_{M_1M_2M_3},\,t^{N_1N_2N_3}\right]&=&18\,\delta^{[N_1N_2}_{[M_1M_2}\,t_{M_3]}{}^{N_3]}-\frac{24}{7}\,\delta^{N_1N_2
N_3 }_{M_1M_2 M_3}\,t\,,\nonumber
\end{eqnarray}
where $t\equiv t_M{}^M$.\par The flux algebra on a twisted torus
is given by a 28--dimensional Lie algebra obtained as follows:
\begin{eqnarray}
\left[Z_M,\,Z_N\right]&=&\alpha\,\tau_{MN}{}^P\,Z_P+\beta\,g_{MNPQ}\,W^{PQ}+\rho\, \tilde{g}\,W_{MN}\,,\nonumber\\
\left[Z_M,\,W^{PQ}\right]&=&\gamma\,\tau_{MR}{}^{[P}\,W^{Q]R}+\sigma\,g_{MM_1M_2M_3}\,\epsilon^{M_1M_2M_3PQRS}\,W_{RS}\,,\nonumber\\
\left[Z_M,\,W_{PQ}\right]&=&
\delta\,\tau_{PQ}{}^L\,W_{ML}\,\nonumber\\
\left[W^{IJ},\,W^{KL}\right]&=&-\frac{\lambda}{2}\,
\tau_{I_1I_2}{}^{[K}\,W_{I_3I_4}\epsilon^{L]IJI_1\dots
I_4}\,,\nonumber\\
\left[W^{IJ},\,W_{KL}\right]&=&\left[W_{IJ},\,W_{KL}\right]=0\,,\nonumber
\end{eqnarray}
where $g_{IJKL},\,\tau_{IJ}{}^K$ satisfy the constraints discussed
in the introduction and the gauge generators read:
\begin{eqnarray}
Z_M&=&
\theta_{M,\,M_1M_2M_3}\,t^{M_1M_2M_3}+\theta_{M,\,N}{}^P\,t_{P}{}^N+\theta_{M,}{}^N\,t_N=
a_1\,g_{M M_1M_2M_3}\,t^{M_1M_2M_3}+\nonumber\\&&a_2
\tau_{MN}{}^P\,t_{P}{}^N+a_3\,\tilde{g}\,t_M\,,\nonumber\\
W^{MN}&=&\theta^{MN,}{}_{PQR}\,t^{PQR}+
\theta^{MN,\,P}\,t_P=\,b_1\,
\tau_{PQ}^{[M}\,t^{N]PQ}+b_2\,\epsilon^{MNM_1\dots M_4
P}\,g_{M_1\dots M_4 }\,t_P\,,\nonumber\\
W_{MN}&=&\theta_{MN,}{}^P\,t_P=c_1\,\tau_{MN}{}^P\,t_P\,.\nonumber
\end{eqnarray}
The various coefficients entering the above formulas are bound to
satisfy the following relations:
 \begin{eqnarray}
a_2&=&\alpha=\frac{\gamma}{2}\,\,\,;\,\,\,a_1=\frac{\beta\,b_1}{3\,a_2}\,\,\,;\,\,\,\,\frac{b_2}{b_1}=\frac{1}{4}\,\frac{a_1}{a_2}
\,,\nonumber\\
c_1\,\sigma
&=&-2\,a_2\,b_2\,\,\,;\,\,\,\frac{\lambda}{\sigma}=\frac{6\alpha}{\beta}\,\,\,;\,\,\,
\delta=\alpha\,\,\,;\,\,\,a_3=\frac{c_1}{a_2}\,\rho\,.\nonumber
\end{eqnarray}
Note that the gauge generators $W^{MN},\,W_{MN}$, as combinations
of $\rE_{7(7)}$ generators, are not linearly independent, but
satisfy the following constraints \cite{dft2}:
\begin{eqnarray}
\tau_{[PQ}}{}^N\,W_{R]N&=&0\,\nonumber\\
b_2\,\epsilon^{M_1M_2M_3M_4PQR}\,g_{M_1M_2M_3M_4}W_{QR}&=&c_1\,\tau_{ST}{}^PW^{ST}\,,\nonumber
\end{eqnarray}
which ensures that at most 21 of them are independent, and thus
that at most 28 vector fields (including the seven vectors
$A^I_\mu$) are involved in the minimal couplings.


\end{document}